\documentclass[english,american]{iopart}
\usepackage[T1]{fontenc}
\usepackage[latin9]{inputenc}
\usepackage{babel}
\usepackage{prettyref}
\usepackage{amstext}
\usepackage{graphicx}
\usepackage[unicode=true,pdfusetitle,
 bookmarks=true,bookmarksnumbered=false,bookmarksopen=false,
 breaklinks=false,pdfborder={0 0 1},backref=false,colorlinks=false]
 {hyperref}
\hypersetup{
 colorlinks=true,linkcolor=blue,citecolor=blue,urlcolor=blue,filecolor=blue}

\makeatletter
\usepackage{iopams}
\usepackage{setstack}

\usepackage[numbers, sort&compress]{natbib}
\usepackage{iopams}
\usepackage{bbold}

\usepackage{stackrel}

\newrefformat{cap}{\hyperref[#1]{Figure~\ref{#1}}}
\newrefformat{fig}{\hyperref[#1]{Figure~\ref{#1}}}
\newrefformat{tab}{\hyperref[#1]{Table ~\ref{#1}}}
\newrefformat{sec}{\hyperref[#1]{Section~\ref{#1}}}
\newrefformat{sub}{\hyperref[#1]{Section~\ref{#1}}}
\newrefformat{cha}{\hyperref[#1]{Chapter~\ref{#1}}}
\newrefformat{app}{\hyperref[#1]{Appendix~\ref{#1}}}

\makeatletter
\newsavebox{\@brx}
\newcommand{\llangle}[1][]{\savebox{\@brx}{\(\m@th{#1\langle}\)}%
  \mathopen{\copy\@brx\kern-0.5\wd\@brx\usebox{\@brx}}}
\newcommand{\rrangle}[1][]{\savebox{\@brx}{\(\m@th{#1\rangle}\)}%
  \mathclose{\copy\@brx\kern-0.5\wd\@brx\usebox{\@brx}}}
\makeatother

\makeatother

\begin{document}
\selectlanguage{english}%
\global\long\def\D{\mathcal{D}}%
\global\long\def\T{\mathrm{T}}%
\global\long\def\Gammafl{\Gamma_{\mathrm{fl},\lambda}}%
\global\long\def\o{\mathcal{O}}%
\global\long\def\tr{\mathrm{tr}}%
\global\long\def\tx{\tilde{x}}%
\global\long\def\tz{\tilde{z}}%
\global\long\def\Ei{\mathrm{Ei}}%
\global\long\def\Sint{S_{\mathrm{int}}}%
\global\long\def\tauM{\tau_{m}}%
\global\long\def\taus{\tau_{s}}%
\global\long\def\tX{\tilde{X}}%
\global\long\def\N{\mathcal{N}}%
\global\long\def\tx{\tilde{x}}%
\global\long\def\tj{\tilde{j}}%
\global\long\def\n{\mathbf{n}}%
\global\long\def\erfc{\mathrm{erfc}}%
\global\long\def\x{\mathbf{x}}%
\foreignlanguage{american}{}
\global\long\def\bj{\mathbf{j}}%
\foreignlanguage{american}{}
\global\long\def\tJ{\tilde{J}}%
\foreignlanguage{american}{}
\global\long\def\by{\mathrm{\mathbf{y}}}%
\foreignlanguage{american}{}
\global\long\def\bx{\mathbf{x}}%
\foreignlanguage{american}{}
\global\long\def\J{\mathbf{J}}%
\foreignlanguage{american}{}
\global\long\def\T{\mathrm{T}}%
\foreignlanguage{american}{}
\global\long\def\th{\tilde{h}}%
\selectlanguage{american}%

\global\long\def\Z{\mathcal{Z}}%
\global\long\def\dd{\mathrm{d}}%
\global\long\def\DD{\mathrm{D}}%
\global\long\def\id{\mathbf{1}}%
\global\long\def\transp{\mathrm{T}}%
\global\long\def\tx{\tilde{x}}%
\global\long\def\coloneq{:=}%
\global\long\def\erfc{\mathrm{erfc}}%
\global\long\def\red{\mathrm{erfc}}%
\global\long\def\bR{\mathbb{R}}%

\title{Capacity of the covariance perceptron}
\author{David Dahmen}
\address{Institute of Neuroscience and Medicine (INM-6) and Institute for Advanced
Simulation (IAS-6) and JARA Institut Brain Structure-Function Relationships
(INM-10), Jülich Research Centre, Jülich, Germany}
\author{Matthieu Gilson{*}}
\address{Center for Brain and Cognition, Universitat Pompeu Fabra, Barcelona,
Spain}
\author{Moritz Helias{*}}
\address{Institute of Neuroscience and Medicine (INM-6) and Institute for Advanced
Simulation (IAS-6) and JARA Institut Brain Structure-Function Relationships
(INM-10), Jülich Research Centre, Jülich, Germany}
\address{Department of Physics, Faculty 1, RWTH Aachen University, Aachen,
Germany}
\begin{abstract}
The classical perceptron is a simple neural network that performs
a binary classification by a linear mapping between static inputs
and outputs and application of a threshold. For small inputs, neural
networks in a stationary state also perform an effectively linear
input-output transformation, but of an entire time series. Choosing
the temporal mean of the time series as the feature for classification,
the linear transformation of the network with subsequent thresholding
is equivalent to the classical perceptron. Here we show that choosing
covariances of time series as the feature for classification maps
the neural network to what we call a 'covariance perceptron'; a mapping
between covariances that is bilinear in terms of weights. By extending
Gardner's theory of connections to this bilinear problem, using a
replica symmetric mean-field theory, we compute the pattern and information
capacities of the covariance perceptron in the infinite-size limit.
Closed-form expressions reveal superior pattern capacity in the binary
classification task compared to the classical perceptron in the case
of a high-dimensional input and low-dimensional output. For less convergent
networks, the mean perceptron classifies a larger number of stimuli.
However, since covariances span a much larger input and output space
than means, the amount of stored information in the covariance perceptron
exceeds the classical counterpart. For strongly convergent connectivity
it is superior by a factor equal to the number of input neurons. Theoretical
calculations are validated numerically for finite size systems using
a gradient-based optimization of a soft-margin, as well as numerical
solvers for the NP hard quadratically constrained quadratic programming
problem, to which training can be mapped.
\end{abstract}
\maketitle

\section{Introduction\label{sec:Introduction-1}}

Binary classification is one of the standard tasks in machine learning.
The simplest artificial neural network that implements classification
is the classical perceptron \cite{Rosenblatt58,Minsky69,Widrow60_96}.
It is a feed-forward mapping from neurons in an input layer to those
of an output layer. Neurons in the output layer receive a weighted
sum of signals from the input layer, which is passed through a Heaviside
nonlinearity, implementing a decision threshold. Inputs to the perceptron,
i.e. activation levels of the input neurons, are specific features
of some data, in the simplest case the pixels of an image or the time
points of a temporal signal. Classification is achieved by training
the weights from the input to the output layer such that the projections
of different input patterns become linearly separable.

Classification is also one essential task for biological neuronal
networks in the brain. However, neurons do not receive different static
features of some input signal, but sequences of action potentials,
so-called spikes, from other neurons. Biological neural networks thus
have to extract the relevant features from these temporal sequences.

To understand biological information processing we need to ask which
features of the temporal sequences contain the relevant information.
Already in the 1950s it has been shown that the firing rate, i.e.
the mean number of spikes in a given time interval, contains information
for example on the presence of a certain feature of a stimulus, such
as the orientation of a bar \cite{Hubel59}. However, cortex also
shows large temporal variability in responses even to the same stimuli
\cite{Arieli96} which raises the question whether this variability
has some functional meaning. Evidence has accumulated that the temporal
fluctuations of neuronal signals around their mean contain behaviorally
relevant information; this holds even to the level of the exact timing
of spikes \cite{Riehle97_1950}. For example, correlations in spike
times across different neurons, such as precise synchronization, have
been shown to code expectations of the animals \cite{Kilavik09_12653}.

Like in artificial neural networks, one mechanism for learning in
the brain is implemented by changing the strengths of connections
between neurons, known as synaptic plasticity. In the middle of the
last century, Donald Hebb postulated the principle 'Cells that fire
together, wire together' \cite{Hebb49,Hertz91}, a plasticity rule
that depends on the activity of the connected neurons. Importantly,
in Hebbian plasticity it is not the overall level of activation of
the neurons (their overall firing rate) that is relevant, but the
temporal coordination of activities. Since then, several learning
rules were proposed to extract information from coordinated fluctuations
\cite{Oja82_267,Barak06_2343}. The model called spike-timing-dependent
plasticity (STDP) \cite{Gerstner96} predicted that changes in synaptic
strengths are sensitive to even the exact spike timings of the pre-
and postsynaptic cells, which was confirmed later on in experiments
\cite{Markram97a,Bi98}. This suggests that temporal fluctuations
in neural activities and their coordination are naturally distinguished
features that shape learning and thereby the function of the neural
circuit.

The simplest measure for coordination between temporal fluctuations
are pairwise covariances between neural activities. In a network of
$m$ neurons, pairwise cross-covariances form an $m(m-1)/2$-dimensional
space for coding information, which is a much larger space than the
space of mean activities. This intuitively suggests that it might
be beneficial for a neural system to make use of the covariances to
represent and process relevant information. In this study, using methods
from statistical mechanics, we examine the performance of such a computational
paradigm which processes information that is represented in fluctuations:
the covariance perceptron \cite{Gilson19_562546}.

Neurons are highly nonlinear processing units. Nevertheless, in large
networks where individual inputs to the neurons are small with respect
to their total input, temporal fluctuations around some stationary
state can be well explained by means of linear response theory \cite{Lindner05_061919}.
In this setting, the network effectively performs a linear transformation
between its inputs and outputs. Linear response theory has been shown
to faithfully capture the statistics of fluctuations in asynchronous,
irregular network states that are observed in cortex \cite{Grytskyy13_131,Dahmen19_inpress}.

In this study, we make use of linear response theory to show that
a network that transforms temporal input signals into temporal output
signals acts as a classical perceptron if the temporal mean of the
network output trajectories is chosen as the relevant feature for
a following binary classification. Choosing, instead, covariances
as the relevant feature of the temporal signals, the same network
acts as a covariance perceptron \cite{Gilson19_562546}, which is
based on a bilinear mapping in terms of weights, but linear from input
to output covariances. The classical perceptron has been studied in
the 1980s in terms of its performance for classification \cite{Gardner88_257}.
Two performance measures are the pattern capacity, the number of patterns
that can be correctly classified into two distinct classes, and the
information capacity, which is the number of bits a conventional computer
requires to realize the same functionality as the neuronal implementation.

Here we set out to study the performance of the covariance-based classification
paradigm by calculating the pattern capacity and information capacity
of the covariance perceptron. Following Gardner's theory of connections,
we use replica symmetric mean-field theory with a saddle-point approximation
to study the volume of possible weight configurations for the classification
problem. The limiting capacity is given by the maximum number of correctly
classifiable stimuli. This analysis shows that the covariance perceptron
can store up to twice as many patterns as the classical perceptron.
Moreover, in case of strongly convergent connectivity, the information
capacity in bits largely exceeds the traditional paradigm by up to
a factor $2m$, with $m$ the number of input neurons. Our work thus
provides evidence that covariance-based information processing in
a biological context can reach superior performance compared to paradigms
that have so far been used in artificial neuronal networks.

\section{Model\label{sec:Model}}

\begin{figure}
\begin{centering}
\includegraphics[width=1\linewidth]{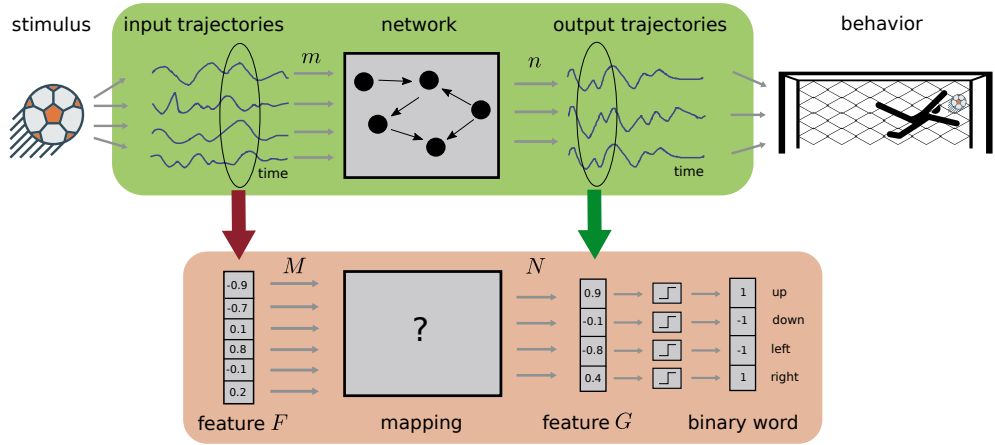}
\par\end{centering}
\caption{\textbf{Setup.} Top: Biological information processing. Sensory information
enters the system as a time series. The dynamics of the neuronal network
produces an outgoing time series of activities. Ultimately, a set
of effectors, for example muscles, are controlled by the resulting
signal. Bottom: Machine-learning view of the biological system. A
set of features $F$ is extracted from the incoming time series. These
are transformed by a mapping to another set of features $G$, each
of which is thresholded to produce a binary output, in the example
encoding the movement direction. The input-output mapping is defined
implicitly by the biological system.\label{fig:Setup}}
\end{figure}

In this study, we consider neural networks that transform patterns
$x(t)$ of $m$ input trajectories $x_{k}(t)$ into patterns $y(t)$
of $n$ output trajectories $y_{i}(t)$ (\prettyref{fig:Setup}).
Using this network transformation and a following hard decision threshold
on the outputs, we want to perform a binary classification of patterns
$x(t)$ into classes labeled by binary words $\zeta$ which entries
can take the values $+1$ and $-1$. Patterns $x(t)$ shall be defined
by an $M$-dimensional feature $F\in\mathbb{R}^{M}$, and classification
shall be performed on an $N$-dimensional feature $G\in\mathbb{R}^{N}$
of the output patterns $y(t)$. Possible features could, for example,
be the signals at a given time point, their temporal average or some
higher order statistics. By transforming patterns $x(t)$ into patterns
$y(t)$, the network maps the feature $F$ of the inputs to the feature
$G$ of the outputs. The connection weights of the network thereby
shall be trained to optimize the classification based on $G$. In
general, features $F$ and $G$ can describe very different characteristics
of the patterns $x(t)$ and $y(t)$, respectively. However, in this
study we choose the case where $F$ and $G$ are of the same type,
which is important for the consideration of multilayer networks.

\begin{figure}
\centering{}\includegraphics[width=1\linewidth]{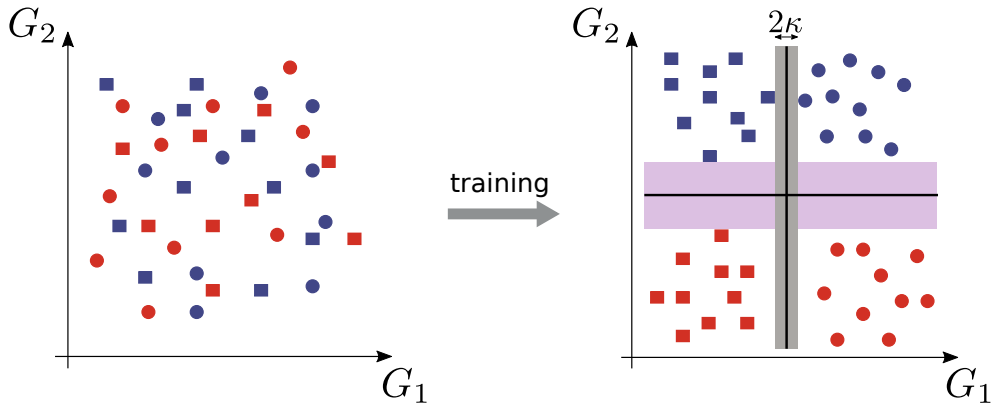}\caption{\textbf{Training.} Classification based on two-dimensional feature
$(G_{1},G_{2})$, here represented by shape (disks/squares) and color
(red/blue). Left: Before training, patterns are scattered randomly
in the space spanned by coordinates $G_{1}$ and $G_{2}$. Right:
After training, weights of the neural network are adjusted such that
patterns with different features become linearly separable with thresholds
(black lines) parallel to the coordinate axes. The margin $\kappa$
(gray area) quantifies the degree of separability: It is the minimal
distance over all data samples to any of the thresholds.\label{fig:margin}}
\end{figure}

An important measure for the quality of the classification is the
margin defined as 
\begin{equation}
\kappa=\min_{1\leq s\leq N}\min_{1\leq r\leq p}\left(\zeta_{s}^{r}G_{s}^{r}\right)\ ,\label{eq:min_margin}
\end{equation}
where $r$ indexes the patterns and $s$ the dimensions of the feature
$G$. The margin measures the smallest distance over all elements
of $G$ from the threshold, here set to $0$. It plays an important
role for the robustness of the classification~\cite{Cortes95_273},
as a larger margin tolerates more noise in the input pattern before
classification is compromised. The margin of the classification is
illustrated in Fig.~\ref{fig:margin}, where each symbol represents
one of the $p$ patterns and the colors and markers indicate the corresponding
category $\zeta^{r}$. Classification is achieved by training the
connection weights of the network to maximize $\kappa$. This optimization
increases the gap and thus the separability between red and blue symbols
and the disks and squares in Fig.~\ref{fig:margin}.

The classification scheme reaches its limit for a certain pattern
load $p$ at which the margin $\kappa$ vanishes. More generally,
one can ask how many patterns the scheme can discriminate while maintaining
a given minimal margin $\kappa$. This defines the pattern capacity
\begin{equation}
\mathcal{P}(\kappa)=\max(p|\kappa).\label{eq:capacity}
\end{equation}
The information capacity is the number of bits required in a conventional
computer, if it were to realize the same classification of the $\mathcal{P}(\kappa)$
patterns
\begin{equation}
\mathcal{I}(\kappa)=\mathcal{P}(\kappa)\left(\log_{2}K+\log_{2}L\right).\label{eq:info_density}
\end{equation}
Here $K$ denotes the number of possible configurations of a single
pattern in the input, and $L$ denotes the number of possible binary
words assigned to the output.

For any general network, one can write the output $y(t)$ as a Volterra
series of the input $x(t)$. Ongoing activity in cortical networks
in many cases shows weakly-fluctuating activity with low correlations.
It has been shown that such small fluctuations around some stationary
state can be well described using linear response theory \cite{Lindner05_061919,Pernice11_e1002059,Trousdale12_e1002408,Grytskyy13_131},
which amounts to a truncation of the Volterra series after the first
order. In such a scenario, the network transformation of small inputs
follows the general form
\begin{equation}
y(t)=\int dt^{\prime}W(t-t^{\prime})x(t^{\prime}),\label{eq:network}
\end{equation}
with some generic linear response kernel $W(t)\in\mathbb{R}^{n\times m}$.
In the following, we will restrict our analysis to this scenario and
study the computational properties of networks operating in the linear
regime.

After having clarified the setup, let us now turn to two specific
examples of features for classification.

\begin{figure}
\begin{centering}
\includegraphics[width=1\linewidth]{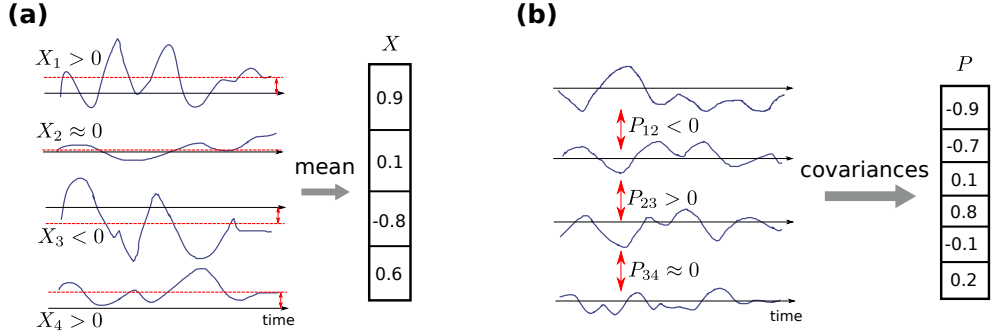}
\par\end{centering}
\caption{\textbf{Extraction of features from time series.} (a) Mean encoding.
Each of the $m$ channels is given by a time series. The temporal
average, the mean $X$, is the feature to be extracted from each time
trace. This yields a vector with one entry per time trace. (b) Covariance
encoding. Representation of information by the covariance between
$m$ time series. These represent $m(m-1)/2$ cross-covariances $P$;
positive covariance corresponds to fluctuations in both time series
that go into the same direction; negative covariance is caused by
antagonistic fluctuations between the two series.\protect \\
\label{fig:feature_extraction}}
\end{figure}

\textbf{Scenario 1: Classical perceptron}

Let's assume that the relevant feature of input trajectories $x_{k}(t)$
is their temporal mean which we here define as $X_{k}=\int dt\,x_{k}(t)$
(Fig. \ref{fig:feature_extraction}a)\footnote{Note that, throughout, we consider observation times $T$ much larger
than the decay time of the linear response kernels $W(t)$. For brevity,
we also drop the trivial normalization by the duration $T$.}. In this case, by integrating Eq.~(\ref{eq:network}) over time,
we see that the network performs the mapping
\begin{equation}
Y=WX
\end{equation}
with weights $W_{ik}$ that are given by the area under the linear
response kernels $W_{ik}=\int dt\,W_{ik}(t).$ The dimension of the
input feature $F$ and output feature $G$ is then $M=m$ and $N=n$,
respectively.

Together with the application of a hard decision threshold on $Y$,
this network transformation acts as $n$ classical perceptrons. The
weights $W_{ik}$ can be trained to reach optimal classification performance.
The pattern capacity of a single classical perceptron classifying
binary patterns has been shown to be \cite{Gardner88_257,Hertz91}
\begin{eqnarray}
\mathcal{P}^{\mathrm{class}}(\kappa) & = & m\left(\int_{-\kappa}^{\infty}\frac{dt}{\sqrt{2\pi}}e^{-t^{2}/2}\left(t+\kappa\right)^{2}\right)^{-1}.\label{pattern_class}
\end{eqnarray}
Note that this capacity does not increase by having more than $n=1$
outputs: all perceptrons receive the same patterns as inputs and therefore
perform the same task.  The information capacity for a classification
scheme based on temporal means follows from Eqs.~(\ref{eq:capacity})
and (\ref{eq:info_density}) with $K=2^{m}$ and $L=2^{n}$ as
\begin{eqnarray}
\mathcal{I^{\mathrm{class}}}(\kappa) & = & \mathcal{P}^{\mathrm{class}}(\kappa)(m+n).\label{eq:info_class}
\end{eqnarray}
Note that in practical applications, one cannot observe input and
output trajectories for infinite time, but only for a finite duration
$T$. The estimate of the mean activity from that finite period $T$
differs from the true temporal mean. Therefore, temporal fluctuations
in the input trajectories constitute a source of noise in this classification
paradigm.

\textbf{Scenario 2: Covariance perceptron}

We now turn to the contrary case where temporal fluctuations, quantified
by the cross-covariances $P_{ij}(\tau)$ of the input trajectories,
shall be the relevant feature for classification (Fig. \ref{fig:feature_extraction}b).
In this case, the feature dimensions are $M=m(m-1)/2$ and $N=n(n-1)/2$.
The mapping from input covariances $P_{ij}(\tau)$ to output covariances
$Q_{ij}(\tau)$ can be derived from Eq.~(\ref{eq:network}) as
\begin{eqnarray*}
Q(\tau) & = & \int dt\,y(t)y^{\mathrm{T}}(t+\tau)-\int dt\,y(t)\int dty^{\mathrm{T}}(t+\tau)\\
 & = & \int ds\int ds^{\prime}\,W(s)P(\tau+s-s^{\prime})W^{\mathrm{T}}(s^{\prime}).
\end{eqnarray*}
The network linearly filters the input covariances $P_{ij}(\tau)$.
If we consider covariances $Q_{ij}=\int d\tau\,Q_{ij}(\tau)$ integrated
across all time lags, we obtain the simple mapping
\begin{equation}
Q=WPW^{\mathrm{T}},\label{eq:mapping}
\end{equation}
which is linear in covariances but bilinear in the weights. Note that
alternatively one could consider a single frequency component $\hat{Q}_{ij}=\int d\tau\,Q_{ij}(\tau)e^{-i\omega\tau}$
and derive an analogous mapping $\hat{Q}=\hat{W}\hat{P}\hat{W}^{\dagger}.$

Therefore, covariance-based classification amounts to a bilinear problem
unlike the classical perceptron, which involves a linear mapping in
terms of weights. In the following, we want to study the capacity
of the covariance perceptron. But before we do so, it is important
to note that the here considered network only acts as if it were a
classical or covariance perceptron. The biological network itself
in both cases receives the full input trajectories and creates the
full output trajectories. However, the mapping between trajectories
can either be optimized for a following classification of the temporal
mean of the output trajectories (classical perceptron) or a classification
based on the covariances of the output trajectories (covariance perceptron).
In this way, the setup here is clearly different from standard machine
learning approaches where one applies a feature selection on the inputs
as a preprocessing step and only classifies this feature vector by
a perceptron. The latter approach amounts to a linear mapping from
$F$ to $G$ rather than a mapping between the entire trajectories
(\prettyref{fig:Setup}).

\section{Theoretical predictions for classification performance based on output
cross-covariances\label{sec:Theory}}

We now derive a theory for the pattern and information capacity of
the covariance perceptron that is exact in the limit of large networks
$m\to\infty$ and that can be compared to the seminal theory by Gardner~\cite{Gardner88_257}
on the capacity of the classical perceptron. Formalizing the classification
problem for a load of $p$ patterns, for each element $Q_{ij}^{r}$
of the readout matrix we draw a random label $\zeta_{ij}^{r}\in\{-1,1\}$
independently for each input pattern $P^{r}$ with $1\leq r\leq p$.
The task of the perceptron is to find a suitable weight matrix $W$
that leads to correct classification for all $p$ patterns. This requirement
reads, for a given margin $\kappa>0$,
\begin{equation}
\zeta_{ij}^{r}Q_{ij}^{r}=\zeta_{ij}^{r}\;\big(WP^{r}W^{\transp}\big)_{ij}>\kappa\ ,\quad\forall\ 1\leq r\leq p\ ,1\leq i<j\leq n.\label{eq:task}
\end{equation}

We assume the patterns $P^{r}$ to be drawn randomly. This random
ensemble allows us to employ methods from disordered systems~\cite{Fischer91}.
Closely following the computation for the classical perceptron by
Gardner~\cite{Gardner88_257,Hertz91}, the idea is to consider the
replication of several covariance perceptrons. The replica, indexed
by $\alpha$ and $\beta$, have the same task defined by Eq.~(\ref{eq:task}).
The sets of patterns $P^{r}$ and labels $\zeta^{r}$ are hence the
same for all replica, but each replicon has its own readout matrix
$W^{\alpha}$. If the task is not too hard, meaning that the pattern
load $p$ is small compared to the number of free parameters $W_{ik}^{\alpha}$,
there are many solutions to Eq.~(\ref{eq:task}). One thus considers
the ensemble of all solutions and computes the typical overlap $R_{ij}^{\alpha\beta}\equiv\sum_{k=1}^{m}W_{ik}^{\alpha}W_{jk}^{\beta}$
between the solution $W^{\alpha}$ and $W^{\beta}$ in two different
replica. The overlap is the scalar product of the weight vectors of
outputs $i$ and $j$. At a certain load $p=\mathcal{P}(\kappa)$,
up to the intrinsic reflection symmetry $W\mapsto-W$ in Eq.~(\ref{eq:task}),
there should only be a single solution left \textemdash the overlap
$R_{ii}^{\alpha\beta}$ between solutions for identical readouts $i=j$,
but in different replica $\alpha\neq\beta$, becomes unity. This pattern
load defines the limiting capacity $\mathcal{P}$.

Technically, the computation proceeds by defining the volume of all
solutions for the whole set of cross-covariances $Q_{ij}^{r}$ that
fulfill the classification task
\begin{equation}
\mathcal{V}=\int\dd W\,\prod_{r}^{p}\prod_{i<j}^{n}\theta\Big(\zeta_{ij}^{r}\;\big(W\,P^{r}\,W^{\transp}\big)_{ij}-\kappa\Big)\ .\label{eq:volume_solutions_main}
\end{equation}
Here $\theta$ denotes the Heaviside function and $\int dW=\prod_{i}^{n}\prod_{k}^{m}\int dW_{ik}$.
This equation is the analogue to Gardner's approach of the perceptron;
see~\cite[Section 10.2, eq. 10.83]{Hertz91}. The typical behavior
of the system for large $m$ is obtained by first taking the average
of $\ln(\mathcal{V})$ over the ensemble of the patterns and labels.
The assumption is that the system is self-averaging; for large $m$
the capacity should not depend much on the particular realization
of patterns. The average of $\ln(\mathcal{V})$ can be computed by
the replica trick $\langle\ln(\mathcal{V})\rangle=\lim_{q\to0}\,\big(\langle\mathcal{V}^{q}\rangle-1\big)/q$~\cite{Fischer91}
which amounts to considering $q$ systems that have identical realizations
of patterns.

\subsubsection{Patterns and classification labels}

In order to perform the average over the patterns and labels, we need
to specify their statistics. Here, we choose a symmetric setting with
independent and identically distributed labels $\zeta_{i<j}^{r}=\pm1$,
each with probability $1/2$. In contrast to inputs to the classical
perceptron which can be chosen independently, covariances involve
constraints related to the fact that a covariance matrix is positive
semidefinite, which implies in particular 
\begin{eqnarray}
P_{kl} & = & P_{lk}\ ,\label{eq:constraints_P0}\\
P_{kk} & \geq & 0\ ,\nonumber \\
|P_{kl}| & \leq & \sqrt{P_{kk}\,P_{ll}}\ ,\nonumber 
\end{eqnarray}
for all indices $k$ and $l$. A simple way to ensure positive semidefiniteness
is to choose each input covariance pattern to be of the form 
\begin{equation}
P^{r}=1_{m}+\chi^{r}\ ,\label{eq:ensemble_input_cov}
\end{equation}
with $1_{m}$ being the $m\times m$ identity matrix and a symmetric
random matrix $\chi^{r}=\left(\chi^{r}\right)^{\transp}$ with vanishing
diagonal entries $\chi_{kk}^{r}=0$ and independent and identically
distributed lower off-diagonal elements $\chi_{k<l}^{r}=\pm c$, each
with probability $f/2$, and $\chi_{k<l}^{r}=0$ with probability
$1-f$.

Here $f$ controls the sparseness (or density) of the non-zero cross-covariances.
The constraint of $P_{kk}=1$ firstly enforces that all information
of the patterns is in the off-diagonal elements. Secondly, it ensures
the positive semidefiniteness for a sufficiently broad range of values
for $c$ and $f$.

The specific form of the input covariances (\ref{eq:ensemble_input_cov})
implies with Eq.~(\ref{eq:mapping}) that 
\[
Q_{ii}^{r}=\sum_{k}^{m}(W_{ik})^{2}+\sum_{k\neq l}^{m}W_{ik}W_{il}\chi_{kl}^{r}\qquad\forall\,i,r\ ,
\]
which, on expectation over patterns, yields 
\begin{equation}
\langle Q_{ii}^{r}\rangle_{\chi}=\sum_{k}^{m}(W_{ik})^{2}\stackrel{!}{=}1\qquad\forall\,i\ .\label{eq:unit_var_output}
\end{equation}
The magnitude of output variances is therefore given by the normalization
of the row vectors of $W$, which we here set to unity. This ensures
that we have a self-consistent scheme, where information in the input
and the output is both encoded only in cross-covariances. Note that
this self-consistency is only reached on expectation as each entry
$Q_{ii}^{r}$ varies around $1$ according to the magnitude of cross-covariances
$\chi$, which, however, for large networks is much smaller than the
variance \cite{Vreeswijk96,Renart10_587,Tetzlaff12_e1002596}. Forcing
$Q_{ii}^{r}=1$ would unnecessarily complicate the analysis; simulation
results below show equal capacity with or without this hard constraint.
Finally, note that the value of unity for the variance does not restrict
the generality of this analysis as any other value of the variance
will lead to another normalization of the weights $W$, which can
be absorbed in the margin $\kappa$ (cf. Eq.~(\ref{eq:task})).

\subsubsection{Pattern and label average}

Given the constraint on the length of the rows in the weight matrix
$W$, Eq.~(\ref{eq:unit_var_output}), we obtain for the expectation
of the power of the volume in Eq.~(\ref{eq:volume_solutions_main})

\begin{eqnarray}
\langle\mathcal{V}^{q}\rangle & = & \left\langle \prod_{\alpha}^{q}\int dW^{\alpha}\,\prod_{i}^{n}\delta\Big(\sum_{k}^{m}(W_{ik}^{\alpha})^{2}-1\Big)\,\prod_{r}^{p}\,\prod_{i<j}^{n}\,\theta\Big(\tilde{Q}_{ij}^{r\alpha}-\kappa\Big)\right\rangle _{\zeta,\chi}\label{eq:Vn}\\
\tilde{Q}_{ij}^{r\alpha} & = & \zeta_{ij}^{r}(W^{\alpha}P^{r}W^{\alpha\transp})_{ij}\ ,\nonumber 
\end{eqnarray}
which we need to study in the limit $q\rightarrow0$. Note the Dirac
delta distribution $\delta$ that describes the constraint (\ref{eq:unit_var_output})
on the length of the weight vectors. The pattern average amounts to
averaging 
\begin{eqnarray}
\left\langle \prod_{\alpha}^{q}\prod_{r}^{p}\prod_{i<j}^{n}\theta\Big(\tilde{Q}_{ij}^{r\alpha}-\kappa\Big)\right\rangle _{\zeta,\chi} & = & \prod_{r}^{p}\left\langle \prod_{\alpha}^{q}\prod_{i<j}^{n}\theta\Big(\tilde{Q}_{ij}^{r\alpha}-\kappa\Big)\right\rangle _{\zeta,\chi}\label{eq:pattern_average}\\
 & = & \left\langle \prod_{\alpha}^{q}\prod_{i<j}^{n}\theta\Big(\tilde{Q}_{ij}^{r\alpha}-\kappa\Big)\right\rangle _{\zeta,\chi}^{p}\nonumber \\
 & \equiv & \mathcal{G}^{p}\,,\nonumber 
\end{eqnarray}
where we used that patterns and labels are uncorrelated (first line)
and follow the same statistics (second line). We can express $\mathcal{G}$
in terms of cumulants of $\tilde{Q}_{ij}^{r\alpha}$ by rewriting
the Heaviside function 
\begin{eqnarray}
\theta\Big(\tilde{Q}_{ij}^{r\alpha}-\kappa\Big) & = & \int_{\kappa}^{\infty}dx_{ij}^{\alpha}\,\delta\big(\tilde{Q}_{ij}^{r\alpha}-x_{ij}^{\alpha}\big)\nonumber \\
 & = & \int_{\kappa}^{\infty}\dd x_{ij}^{\alpha}\int_{-i\infty}^{i\infty}\frac{\dd\tx_{ij}^{\alpha}}{2\pi i}\,e^{\tx_{ij}^{\alpha}\big(\tilde{Q}_{ij}^{r\alpha}-x_{ij}^{\alpha}\big)},\label{eq:Heaviside}
\end{eqnarray}
such that 
\begin{eqnarray}
\mathcal{G} & = & \prod_{i<j}^{n}\left\{ \int Dx_{ij}\int D\tilde{x}_{ij}\right\} \,e^{-\sum_{\alpha}^{q}\sum_{i<j}^{n}\tx_{ij}^{\alpha}x_{ij}^{\alpha}}\left\langle e^{\sum_{\alpha}^{q}\sum_{i<j}^{n}\tx_{ij}^{\alpha}\tilde{Q}_{ij}^{r\alpha}}\right\rangle _{\zeta,\chi}\nonumber \\
 & = & \prod_{i<j}^{n}\left\{ \int Dx\int D\tilde{x}\,e^{-\sum_{\alpha}^{q}\tx^{\alpha}x^{\alpha}+\frac{1}{2}\sum_{\alpha,\beta}^{q}\,\tx^{\alpha}\left\langle \left\langle \tilde{Q}_{ij}^{r\alpha}\tilde{Q}_{ij}^{r\beta}\right\rangle \right\rangle _{\zeta,\chi}\tx^{\beta}}\right\} \nonumber \\
 & \equiv & \prod_{i<j}^{n}\mathcal{G}_{ij}\,.\label{eq:defG}
\end{eqnarray}
Here we used the abbreviations $\int Dx\equiv\prod_{\alpha}^{q}\int_{\kappa}^{\infty}\dd x^{\alpha}$
and $\int D\tx\equiv\prod_{\alpha}^{q}\int_{-i\infty}^{i\infty}\frac{\dd\tx^{\alpha}}{2\pi i}$,
where indices of integration variables in the second line have been
dropped. We further identified the moment generating function of the
variables $\tilde{Q}_{ij}^{r\alpha}$ with respect to the statistics
of $\zeta$ and $\chi$ (first line). The latter can be expanded in
cumulants (second line): In the large-$m$ limit, this expansion can
be truncated at the second cumulant $\left\langle \left\langle \tilde{Q}_{ij}^{r\alpha}\tilde{Q}_{kl}^{r\beta}\right\rangle \right\rangle _{\zeta,\chi}$
in a similar fashion to the classical perceptron~\cite{Hertz91}.
Due to the independence of labels $\zeta_{ij}$ and $\zeta_{kl}$,
the second cumulants vanish if $i\neq k$ or $j\neq l$. The diagonal
elements are given by 

\begin{eqnarray}
\left\langle \left\langle \tilde{Q}_{ij}^{r\alpha}\tilde{Q}_{ij}^{r\beta}\right\rangle \right\rangle _{\zeta,\chi} & = & \Big(\sum_{k}^{m}W_{ik}^{\alpha}W_{jk}^{\alpha}\Big)\,\Big(\sum_{k}^{m}W_{ik}^{\beta}W_{jk}^{\beta}\Big)\label{eq:sec_cum}\\
 &  & +fc^{2}\,\Big(\sum_{k}^{m}W_{ik}^{\alpha}W_{ik}^{\beta}\Big)\,\Big(\sum_{k}^{m}W_{jk}^{\alpha}W_{jk}^{\beta}\Big)\nonumber \\
 &  & +fc^{2}\,\Big(\sum_{k}^{m}W_{ik}^{\alpha}W_{jk}^{\beta}\Big)\,\Big(\sum_{k}^{m}W_{ik}^{\beta}W_{jk}^{\alpha}\Big)\ .\nonumber 
\end{eqnarray}
The first term arises from the diagonal $1_{m}$ in the patterns $P^{r}$
in Eq.~(\ref{eq:ensemble_input_cov}), and the third term stems from
the symmetry of $P^{r}$. These two terms would be absent for completely
uncorrelated i.i.d. matrices $P^{r}$. In the second and third lines
in Eq.~(\ref{eq:sec_cum}) we added a single term $k=l$ which is
negligible in the large-$m$ limit. We see that the only dependence
on the sparseness $f$ and the magnitude $c$ of input covariances
is in the form $fc^{2}$ \textemdash it does not depend on these two
parameters separately. The problem is, moreover, now symmetric in
all $i<j$ index pairs. We also observe that the product over all
$p$ patterns has identical factors that do not depend on the pattern
index $r$, so that we only get this factor to the $p$-th power.

\subsubsection{Auxiliary field formulation}

Starting from Eq.~(\ref{eq:sec_cum}), we now define the auxiliary
fields as 
\begin{equation}
R_{ij}^{\alpha\beta}\equiv\sum_{k}^{m}W_{ik}^{\alpha}W_{jk}^{\beta}\ ,\label{eq:def_aux_fields-1}
\end{equation}
for $i<j$ and $\alpha\neq\beta$. The field $R_{ij}^{\alpha\alpha}$
for $i\neq j$ measures the overlap between weight vectors to different
units in the same replicon $\alpha$. It contributes to the average
value of $Q_{ij}^{r\alpha}$ because the unit diagonal (common to
all $P^{r}$) is weighted by $R_{ij}^{\alpha\alpha}$. Hence the output
$Q_{ij}^{r\alpha}$ will be displaced by $R_{ij}^{\alpha\alpha}$
irrespective of the realization of $P^{r}$. $R_{ij}^{\alpha\beta}$
for $\alpha\neq\beta$ measures the overlap of weight vectors in different
replica. For $\alpha=\beta$ and $i=j$ we have $R_{ii}^{\alpha\alpha}=1$
due to Eq.~(\ref{eq:unit_var_output}). The definitions Eq. (\ref{eq:def_aux_fields-1})
are enforced by integrations over Dirac distributions
\begin{eqnarray}
\delta\big(\sum_{k}^{m}W_{ik}^{\alpha}W_{jk}^{\beta}-R_{ij}^{\alpha\beta}\big) & = & \int_{-i\infty}^{i\infty}\frac{\dd\tilde{R}_{ij}^{\alpha\beta}}{2\pi i}\,e^{-\tilde{R}_{ij}^{\alpha\beta}\sum_{k}^{m}W_{ik}^{\alpha}W_{jk}^{\beta}+\tilde{R}_{ij}^{\alpha\beta}R_{ij}^{\alpha\beta}}\label{eq:deltas}
\end{eqnarray}
Note that the same Fourier representation can also be used for the
length constraint on the weight vectors $\delta\big((W^{\alpha}W^{\alpha\T})_{ii}-1\big)$.
After inserting auxiliary fields into Eq.~(\ref{eq:defG}), the integration
over weights $W_{ik}^{\alpha}$ only applies to the first term in
Eq.~(\ref{eq:deltas}) for all indices
\begin{eqnarray}
 &  & \prod_{\alpha}^{q}\prod_{i}^{n}\prod_{k}^{m}\int dW_{ik}^{\alpha}\,\exp\big(-\sum_{\alpha,\beta}^{q}\sum_{i\leq j}^{n}\tilde{R}_{ij}^{\alpha\beta}\sum_{k}^{m}W_{ik}^{\alpha}W_{jk}^{\beta}\big)\nonumber \\
 & = & \left(\prod_{\alpha}^{q}\prod_{i}^{n}\int dW_{i}^{\alpha}\,\exp\big(-\sum_{\alpha,\beta}^{q}\sum_{i\leq j}^{n}\tilde{R}_{ij}^{\alpha\beta}W_{i}^{\alpha}W_{j}^{\beta}\big)\right)^{m}\nonumber \\
 & \equiv & \mathcal{F}^{m}\,.\label{eq:defF}
\end{eqnarray}
Here, we used that the expression factorizes in the index $k$ so
that we get the same integral to the $m$-th power, one factor for
each component $W_{ik}^{\alpha}\,\forall k=1,\ldots,m$, allowing
us to define a single integration variable $W_{i}^{\alpha}$. Gathering
contributions from the second term in Eq.~(\ref{eq:deltas}) for
all indices defines
\begin{equation}
\mathcal{H}\equiv\sum_{\alpha,\beta}^{q}\sum_{i\leq j}^{n}\tilde{R}_{ij}^{\alpha\beta}R_{ij}^{\alpha\beta}.\label{eq:defH}
\end{equation}
Expressing $\left\langle \mathcal{V}^{q}\right\rangle $ in terms
of $\mathcal{F}$, $\mathcal{G}_{ij}$ and $\mathcal{H}$ then yields
\begin{eqnarray}
\left\langle \mathcal{V}^{q}\right\rangle  & = & \int\dd R\int\dd\tilde{R}\,\exp\big(m\,\ln(\mathcal{F})+p\sum_{i<j}^{n}\ln(\mathcal{G}_{ij})+\mathcal{H}\big),\label{eq:V2_before_saddle}
\end{eqnarray}
with $\int\dd R\equiv\prod_{\alpha,\beta}^{q}\prod_{i<j}^{n}\int\dd R_{ij}^{\alpha\beta}\prod_{\alpha\neq\beta}^{q}\int\dd R_{ii}^{\alpha\beta}$
and $\int\dd\tilde{R}=\prod_{\alpha,\beta}^{q}\prod_{i\leq j}^{n}\int_{-i\infty}^{i\infty}\frac{\dd\tilde{R}_{ij}^{\alpha\beta}}{2\pi i}$.

The leading order behavior for $m\to\infty$ follows as a mean-field
approximation in the auxiliary variables $R_{ij}^{\alpha\beta}$.
Therefore, we are interested in the saddle points of the integrals
$\int\dd R\int\dd\tilde{R}$ and search for a replica-symmetric solution.
We therefore set 
\begin{eqnarray}
R_{ij}^{\alpha\alpha}=R_{ij}^{=}, & \hspace{1em}\tilde{R}_{ij}^{\alpha\alpha}=\tilde{R}_{ij}^{=}\\
R_{ij}^{\alpha\beta}=R_{ij}^{\neq}, & \hspace{1em}\tilde{R}_{ij}^{\alpha\beta}=\tilde{R}_{ij}^{\neq}\nonumber 
\end{eqnarray}
for $\alpha\neq\beta$. For replica symmetry, the exponent in $\mathcal{G}_{ij}$
simplifies to 
\begin{eqnarray}
 &  & \sum_{\alpha,\beta}^{q}\,\tx^{\alpha}\Big(R_{ij}^{\alpha\alpha}R_{ij}^{\beta\beta}+fc^{2}\,R_{ii}^{\alpha\beta}\,R_{jj}^{\alpha\beta}+fc^{2}\,R_{ij}^{\alpha\beta}\,R_{ij}^{\beta\alpha}\Big)\tx^{\beta}\nonumber \\
 &  & =(\lambda_{ij}^{=}-\lambda_{ij}^{\neq})\sum_{\alpha}^{q}\,\tx^{\alpha}\tx^{\alpha}+\lambda_{ij}^{\neq}\left(\sum_{\alpha}^{q}\,\tx^{\alpha}\right)^{2}\ ,\label{eq:lambda=00003D}
\end{eqnarray}
with $\lambda_{ij}^{=}=fc^{2}\,R_{ii}^{=}R_{jj}^{=}+(1+fc^{2})\,R_{ij}^{=2}$
and $\lambda_{ij}^{\neq}=fc^{2}\,R_{ii}^{\neq}R_{jj}^{\neq}+R_{ij}^{=2}+fc^{2}\,R_{ij}^{\neq2}$.
The replica are coupled by the factor $\lambda_{ij}^{\neq}$, which
renders $\int\DD\tilde{x}$ in $\mathcal{G}_{ij}$ a $q$-dimensional
integral. In order to apply the limit $q\rightarrow0$, it is convenient
to decouple the replica by performing the Hubbard-Stratonovich transformation
\begin{equation}
e^{\frac{1}{2}\lambda_{ij}^{\neq}\Big(\sum_{\alpha}^{q}\,\tx^{\alpha}\Big)^{2}}=\int Dt\,e^{t\sqrt{\lambda_{ij}^{\neq}}\sum_{\alpha}^{q}\,\tx^{\alpha}}\ ,
\end{equation}
with $\int Dt\equiv\int_{-\infty}^{\infty}\frac{dt}{\sqrt{2\pi}}e^{-t^{2}/2}$,
which turns the $2q$-dimensional integral over $x^{\alpha}$ and
$\tilde{x}^{\alpha}$ for $1\leq\alpha\leq q$ into a Gaussian integral
over the $q$-th power of a function $g_{ij}(t)$ that is given by
a two-dimensional integral 
\begin{eqnarray}
g_{ij}(t) & = & \int_{\kappa}^{\infty}\dd x\int_{-i\infty}^{i\infty}\frac{\dd\tilde{x}}{2\pi i}\,e^{\frac{1}{2}(\lambda_{ij}^{=}-\lambda_{ij}^{\neq})\tilde{x}^{2}+t\sqrt{\lambda_{ij}^{\neq}}\tilde{x}-\tx x}\nonumber \\
 & = & \frac{1}{2}\erfc(a_{ij}(t))\ ,
\end{eqnarray}
with $a_{ij}(t)=(\kappa-t\sqrt{\lambda_{ij}^{\neq}})/\sqrt{2(\lambda_{ij}^{=}-\lambda_{ij}^{\neq})}$.
The resulting form of $\mathcal{G}_{ij}$ allows us to take advantage
of the $q\rightarrow0$ limit by approximating 
\begin{eqnarray}
\ln(\mathcal{G}_{ij}) & = & \ln\left\langle g_{ij}(t)^{q}\right\rangle =\ln\left\langle \exp\left(q\ln\big(g_{ij}(t)\big)\right)\right\rangle \label{eq:lnG}\\
 & \rightarrow & \ln\left(1+q\left\langle \ln\big(g_{ij}(t)\big)\right\rangle \right)\rightarrow q\left\langle \ln\big(g_{ij}(t)\big)\right\rangle \nonumber \\
 & = & q\left\langle \ln\Big(\erfc\big(a_{ij}(t)\big)\Big)\right\rangle +\ln(1/2)\ .\nonumber 
\end{eqnarray}
\begin{figure}
\centering{}\includegraphics{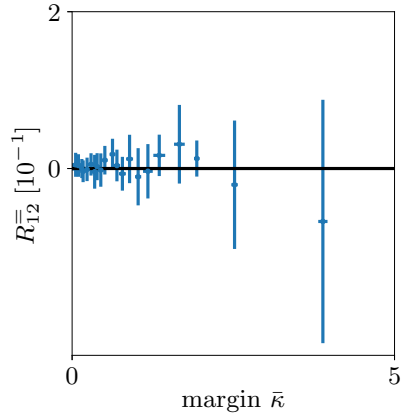}\caption{\textbf{Overlap between weight vectors of different outputs.} Overlap
$R_{12}^{=}=\check{W}^{1\protect\transp}\check{W}^{2}$ between the
pair of row vectors involved in the calculation of the readout covariance
$Q_{12}$. Symbols from numerical optimization (method=IPOPT, see
\prettyref{sec:Results-1}); error bars show standard error from $5$
realizations of patterns and $10$ initial conditions for the numerical
solver; solid black line from theory in the large $m$ limit, which
predicts $R_{12}^{=}\to0$. Other parameters: $m=50$, $n=2$, $f=0.1$,
$c=0.5$. \label{fig:overlap}}
\end{figure}

\subsubsection{Replica limit $q\rightarrow0$\label{subsec:limit}}

So far, we considered $q$ replica of the system, where $q$ was a
natural number. In order to find the typical behavior of $\mathcal{V}$,
the replica trick requires us to study the limit $q\rightarrow0$.
Using $q(q-1)=-q+\mathcal{O}(q^{2})$, we get 
\begin{equation}
\mathcal{H}=q\sum_{i\leq j}^{n}\tilde{R}_{ij}^{=}R_{ij}^{=}-q\sum_{i\leq j}^{n}\tilde{R}_{ij}^{\neq}R_{ij}^{\neq},
\end{equation}
which gives rise to the following saddle point equations 
\begin{eqnarray}
R_{ij}^{=} & = & -\frac{m}{q}\frac{\partial\ln(\mathcal{F})}{\partial\tilde{R}_{ij}^{=}}\ ,\hspace{1em}R_{ij}^{\neq}=\frac{m}{q}\frac{\partial\ln(\mathcal{F})}{\partial\tilde{R}_{ij}^{\neq}}\label{eq:R=00003D00003D}\\
\tilde{R}_{ij}^{=} & = & -\frac{p}{q}\sum_{k<l}^{n}\frac{\partial\ln(\mathcal{G}_{kl})}{\partial R_{ij}^{=}}\ ,\hspace{1em}\tilde{R}_{ij}^{\neq}=\frac{p}{q}\sum_{k<l}^{n}\frac{\partial\ln(\mathcal{G}_{kl})}{\partial R_{ij}^{\neq}}.\label{eq:Rtildeneq}
\end{eqnarray}
The above equations show that we need to find the contribution of
$\ln(\mathcal{F})$ and $\ln(\mathcal{G}_{ij})$ proportional to $q$
as this is the only term surviving in the $q\rightarrow0$ limit.

\subsubsection{Limiting pattern load $p\rightarrow\mathcal{P}(\kappa)$\label{subsec:Limiting-capacity}}

We are interested in the limit $p\rightarrow\mathcal{P}(\kappa)$,
which denotes the point where only a single solution is found: the
overlap $R_{ii}^{\neq}\to R_{ii}^{=}=1$ of the readout between replica
is identical to the length of the vector in each individual replicon,
so only a single solution is found. So we set $R_{ii}^{\neq}=1-\epsilon$
and study the limit $\epsilon\rightarrow0$ for all $i\in[1,m]$ simultaneously.
We need to be careful in taking this limit as $\ln(\mathcal{G}_{ij})$
is singular for $\epsilon=0$. The saddle-point equations relate derivatives
of $\ln(\mathcal{G}_{ij})$ to tilde-fields, which in turn are defined
by $\ln(\mathcal{F})$. A singularity in $\ln(\mathcal{G}_{ij})$
at $\epsilon=0$ therefore implies also a singularity in $\ln(\mathcal{F})$.
These singularities will cancel in the following calculation of the
capacity.

We first focus on the fields $R_{ij}^{=}$ and $R_{ij}^{\neq}$ for
$i<j$: The function $\ln G_{ij}$ depends quadratically on $R_{ij}^{=}$
and $R_{ij}^{\neq}$ (see Eq.~(\ref{eq:lambda=00003D})). The same
is true for $\mathcal{F}$, which can be seen by Taylor expansion
around $\tilde{R}_{ij}^{=}=\tilde{R}_{ij}^{\neq}=0$ (cp. Eq.~(\ref{eq:defF})):
all odd Taylor coefficients vanish since they are determined by odd
moments of a Gaussian integral with zero mean. By rewriting Eq.~(\ref{eq:Rtildeneq})
as $\tilde{R}_{ij}^{=}=-2R_{ij}^{=}\frac{p}{q}\sum_{k<l}^{n}\frac{\partial\ln(\mathcal{G}_{kl})}{\partial R_{ij}^{=2}}$
and $R_{ij}^{=}=-2\tilde{R}_{ij}^{=}\frac{m}{q}\frac{\partial\ln(\mathcal{F})}{\partial\tilde{R}_{ij}^{=2}}$,
respectively, and analogously for $\tilde{R}_{ij}^{\neq}$ and $R_{ij}^{\neq}$,
we see that $R_{ij}^{=}=\tilde{R}_{ij}^{=}=R_{ij}^{\neq}=\tilde{R}_{ij}^{\neq}=0$
is a solution to the saddle point equations. This solution makes sense
as $R_{ij}^{=}$ represents a displacement of the $Q_{ij}$ irrespective
of the label $\zeta^{r}=\pm1$ . Thus the perceptron would lose flexibility
in assigning arbitrary labels to patterns; a non-vanishing value would
therefore hinder the classification. \prettyref{fig:overlap} shows
the numerically calculated overlap $R_{ij}^{=}$ to be close to zero,
as predicted by the theory. At the point of limiting capacity all
replica find the same solution. Therefore, also the overlap $R_{ij}^{\neq}$
across replica must vanish. Using $\tilde{R}_{ij}^{=}=\tilde{R}_{ij}^{\neq}=0$,
an analogous procedure as in Section~\ref{subsec:limit} can be performed
to calculate the term $\ln(\mathcal{F})$ in the $q\rightarrow0$
limit 
\begin{equation}
\ln(\mathcal{F})\rightarrow-\frac{1}{2}q\sum_{i}^{n}\left(\ln\left(\tilde{R}_{ii}^{=}-\tilde{R}_{ii}^{\neq}\right)+\tilde{R}_{ii}^{\neq}/\left(\tilde{R}_{ii}^{=}-\tilde{R}_{ii}^{\neq}\right)\right)+\mathrm{const}\ .\label{eq:lnF}
\end{equation}
Then Eq.~(\ref{eq:R=00003D00003D}) can be easily solved to obtain
\begin{equation}
\tilde{R}_{ii}^{=}=-\frac{m}{2}\frac{1-2\epsilon}{\epsilon^{2}},\hspace{1em}\tilde{R}_{ii}^{\neq}=-\frac{m}{2}\frac{1-\epsilon}{\epsilon^{2}}\ .\label{eq:tildeReq}
\end{equation}
Inserting this solution into Eq.~(\ref{eq:Rtildeneq}) and using
Eq.~(\ref{eq:lnG}), we get in the limit $\epsilon\rightarrow0$
\begin{eqnarray*}
-\frac{m}{2}\frac{1}{\epsilon^{2}} & = & \mathcal{P}(\kappa)\sum_{k<l}^{n}\left(\frac{\partial\left\langle \ln\Big(\erfc(a_{kl}(t))\Big)\right\rangle }{\partial R_{kk}^{\neq}}\delta_{ki}+(k\leftrightarrow l)\right)\\
 & = & \mathcal{P}(\kappa)\sum_{k<l}^{n}\int Dt\frac{\frac{\partial}{\partial a_{kl}(t)}\erfc(a_{kl}(t))}{\erfc(a_{kl}(t))}\left(\frac{\partial a_{kl}(t)}{\partial R_{kk}^{\neq}}\delta_{ki}+\frac{\partial a_{kl}(t)}{\partial R_{ll}^{\neq}}\delta_{li}\right)\\
 & = & \mathcal{P}(\kappa)\sum_{k<l}^{n}\int Dt\frac{-\frac{2}{\sqrt{\pi}}e^{-a_{kl}(t)^{2}}}{\erfc(a_{kl}(t))}\left(\frac{\partial a_{kl}(t)}{\partial R_{kk}^{\neq}}\delta_{ki}+\frac{\partial a_{kl}(t)}{\partial R_{ll}^{\neq}}\delta_{li}\right).
\end{eqnarray*}
For $\epsilon\rightarrow0$ the function $a_{kl}(t)$ goes to negative
infinity for $t>\kappa/\sqrt{fc^{2}}$ and $\erfc(a_{kl}(t))\rightarrow2$.
In this case the numerator in the integrand makes the integral vanish.
Therefore, we can restrict the integration range to $t\in(-\infty,\kappa/\sqrt{fc^{2}}]$,
where $a_{kl}(t)\rightarrow\infty$ for $\epsilon\rightarrow0$, such
that we can insert the limit behavior of $\erfc(a_{kl}(t))\rightarrow e^{-a_{kl}(t)^{2}}/(\sqrt{\pi}a_{kl}(t))$.
Using 
\begin{equation}
\frac{\partial a_{kl}(t)^{2}}{\partial R_{kk}^{\neq}}\rightarrow\frac{\left(\kappa-t\sqrt{fr^{2}}\right)^{2}}{2fr^{2}(1-(1-\epsilon)^{2})^{2}}+\mathcal{O}\left(\frac{1}{\epsilon}\right)\rightarrow\frac{\left(\bar{\kappa}-t\right)^{2}}{8\epsilon^{2}}+\mathcal{O}\left(\frac{1}{\epsilon}\right),\label{eq:limit}
\end{equation}
where we introduced $\bar{\kappa}=\kappa/\sqrt{fr^{2}}$, the limiting
number of patterns $\mathcal{P}(\kappa)$ follows from 
\[
\frac{m}{\epsilon^{2}}=\mathcal{P}(\kappa)\int_{-\infty}^{\bar{\kappa}}\frac{dt}{\sqrt{2\pi}}\exp\left(-\frac{t^{2}}{2}\right)\frac{\left(\bar{\kappa}-t\right)^{2}}{4\epsilon^{2}}\sum_{k<l}^{n}\left(\delta_{ki}+\delta_{li}\right)
\]
as
\begin{equation}
\mathcal{P}(\kappa)=4\frac{m}{n-1}\left(\int_{-\bar{\kappa}}^{\infty}\frac{dt}{\sqrt{2\pi}}\exp\left(-\frac{t^{2}}{2}\right)\left(t+\bar{\kappa}\right)^{2}\right)^{-1}.\label{eq:p_final}
\end{equation}

\section{Results\label{sec:Results-1}}

In the following, we present the key conclusions from the theoretical
derivation of Eq. (\ref{eq:p_final}) as well as numerical validations.

\subsection*{The pattern capacity grows linearly in the number of inputs.}

Analogous to the classical perceptron, the pattern capacity is an
extensive quantity in the number of inputs:~$\mathcal{P\sim}m$.
This result is obvious as a higher-dimensional space facilitates classification.
Formally, this is shown by the problem factorizing in the input indices
in the saddle-point approximation.

\subsection*{The pattern capacity depends on the rescaled margin.}

The pattern capacity only depends on the margin through the parameter
$\bar{\kappa}\equiv\kappa/\sqrt{fc^{2}}$, which measures the margin
$\kappa$ relative to the standard deviation of the readout (\prettyref{fig:pattern_cap}).
This is also true for the classical perceptron, which was originally
analyzed for $fc^{2}=1$. In the latter case, the result is simple
as, for random patterns $\xi$ with standard deviation $\sqrt{fc^{2}}$,
the inequality $\zeta_{i}\big(W\xi\big)_{i}>\kappa$ is equivalent
to $\zeta_{i}\big(W\tilde{\xi}\big)_{i}>\kappa/\sqrt{fc^{2}}=\bar{\kappa}$
with rescaled patterns $\tilde{\xi}$ of unit standard deviation.
In the case of the covariance perceptron, the situation is more subtle
due to the identity matrix contained in the patterns $P=1+\chi$.
However, due to the orthogonality of the weight vectors $WW^{\transp}=0$
(\prettyref{fig:overlap}) the same scaling arguments hold in Eq.~(\ref{eq:task}).
Note that the orthogonality arises naturally as a non-zero overlap
$R_{ij}^{=}$ between different weight vectors ($i\neq j$) would
cause a bias in outputs and therefore hinder classification of $Q_{ij}$
around zero.

Similarly, the replica-symmetric solution is agnostic to the specificity
in patterns $P$ with regard to symmetry of $\chi$. This can be seen
from the terms including $R_{ij}^{\neq}$, which only arise due to
the symmetry: the replica-symmetric solution of the saddle-point equations
implies $R_{ij}^{\neq}=0$, i.e. orthogonality of different weight
vectors in different replica. Physically, it makes sense that at the
limiting pattern load, all replica behave similarly.

In addition to the fact that for both perceptrons $\bar{\kappa}$
is the determining quantity, the dependence of the pattern capacity
on $\bar{\kappa}$ is also identical. The derivation in Section~\ref{sec:Theory}
explains this structural similarity: The functions $\mathcal{F}$
and $\mathcal{H}$ are the same for both perceptrons as they follow
from the length constraint on the weight vectors and the introduction
of the auxiliary fields in Eq. (\ref{eq:def_aux_fields-1}). The latter
are identical for both perceptrons. Also the structure of $\mathcal{G}$
and its dependence on $a(t)$ is the same in both cases, resulting
in the integral in Eq.~(\ref{eq:p_final}).

\subsection*{A single readout has a four times higher pattern capacity than the
classical perceptron.}

For a single readout ($N=1\rightarrow n=2$), the pattern capacity
of the covariance perceptron is four times larger than the pattern
capacity of the classical perceptron (\prettyref{fig:pattern_cap}b)
\begin{eqnarray}
\mathcal{P}^{\mathrm{cov}}(\bar{\kappa}) & = & 4\mathcal{P}^{\mathrm{class}}(\bar{\kappa}).\label{eq:C_cov}
\end{eqnarray}

The superior pattern capacity of the covariance perceptron can be
understood intuitively: For a single readout, the problem to be solved
reads $Q_{12}=\check{W}^{1\transp}\,P\,\check{W}^{2}$, which is bilinear
in $\check{W}^{1}$ and $\check{W}^{2}$, the first and second row
of the weight matrix $W$. Choosing $\check{W}^{1}$ as a random vector
can only lead to the same or worse classification performance than
optimizing $\check{W}^{1}$ for the given set of patterns: the product
$\check{W}^{1\transp}\,P\equiv\xi$ can be considered a random pattern
and optimizing only $\check{W}^{2}$ thus amounts to a classical perceptron.
Therefore, optimizing both, $\check{W}^{1}$ and $\check{W}^{2}$,
the performance must be larger or equal to that of a linear perceptron.
For $N\gg1$, different rows $\check{W}^{i}$ are not independent,
as each row determines multiple output covariances $Q_{ij}$. These
amount to additional constraints in the optimization and thus a reduction
in pattern capacity (see below).

Formally, the different scaling (factor $4$ in Eq.~\ref{eq:p_final})
of the pattern capacity for the two perceptrons comes from the squared
appearance of the auxiliary fields in $\lambda^{=}$ and $\lambda^{\neq}$
(Eq.~\ref{eq:lambda=00003D}), as opposed to a linear fashion for
the case of the classical perceptron. In Eq.~(\ref{eq:limit}), the
leading behavior in the limit $\epsilon\rightarrow0$ is therefore
$(1-(1-\epsilon)^{2})^{2}\rightarrow4\epsilon^{2}+\o(\epsilon^{4})$
rather than $(1-(1-\epsilon))^{2}\rightarrow\epsilon^{2}$ for the
classical perceptron.

\subsection*{A single readout uses less resources to classify the same number
of patterns than the classical perceptron.}

\begin{figure}
\centering{}\includegraphics{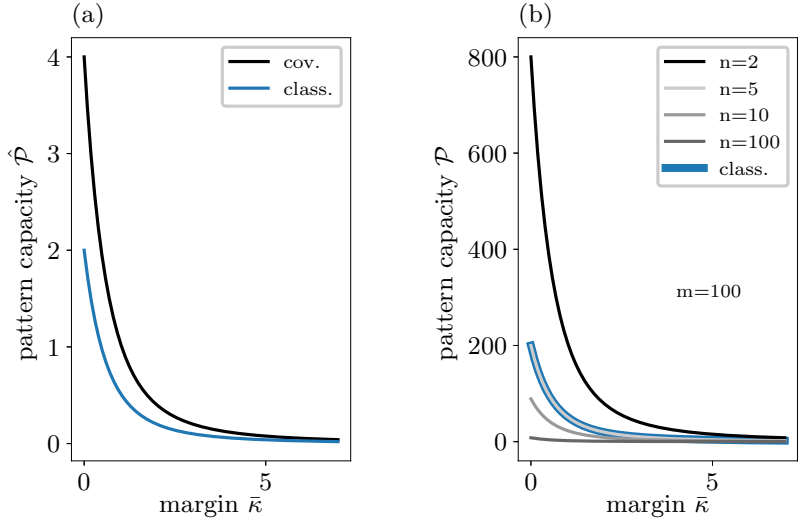}\caption{\textbf{Pattern capacity.} (\textbf{a}) Pattern capacity per synapse
for a single readout as a function of the margin $\bar{\kappa}=\kappa/\sqrt{fc^{2}}$
relative to the typical variance $\sqrt{fc^{2}}$ of an element of
the readout matrix (black: covariance perceptron with $n=2$; blue:
classical perceptron with $n=1$). (\textbf{b}) Pattern capacity of
the covariance perceptron for different numbers of readouts (solid
grey lines) in comparison to classical perceptron (blue dashed line).
Parameters: $f=0.2$, $c=0.5$, $m=100$. \label{fig:pattern_cap}}
\end{figure}

Given the arguments above, the higher pattern capacity of the covariance
perceptron seems natural, as a single readout already implies $n=2$
outputs, i.e. twice as many tunable weights compared to a single classical
perceptron. However, the pattern capacity is not twice as high as
for the classical perceptron, as would be expected from this doubling
of weights. Instead, the joint optimization leads to a synergy effect,
giving rise to an overall difference of a factor $4$ in the pattern
capacity, and a doubling in the pattern capacity per synapse (\prettyref{fig:pattern_cap}a)
\begin{eqnarray}
\hat{\mathcal{P}}^{\mathrm{cov}}(\bar{\kappa}) & =\frac{\mathcal{P}^{\mathrm{cov}}(\bar{\kappa})}{2m}= & 2\frac{\mathcal{P}^{\mathrm{class}}(\bar{\kappa})}{m}=2\hat{\mathcal{P}}^{\mathrm{class}}(\bar{\kappa}).\label{eq:C_cov-1}
\end{eqnarray}
The pattern capacity per synapse is a useful measure to compare the
performance of the two perceptrons, as it accounts for the number
of tunable weights, which in turn determine the complexity of the
computational paradigm.

\subsection*{The pattern capacity decreases with the number of outputs.}

Contrary to classical perceptrons, which have a pattern capacity independent
of $n$, the pattern capacity of the covariance perceptron decreases
with increasing number $n$ of outputs (\prettyref{fig:pattern_cap}b):
$\mathcal{P}^{\mathrm{cov}}\sim(n-1)^{-1}$. This can be understood
as follows: For classical perceptrons, the weights to different readouts
are independent. Therefore, adding more readouts does not impact the
determination of possible weight configurations of the already existing
readouts. The covariance perceptron, however, constitutes a bilinear
mapping in terms of weights which leads to shared weight vectors for
different output covariances. As an example, the weight vector for
neuron $1$ impacts the output $Q_{1j}$ for all outputs $j>1$. Thus,
adding one more output, let's say the $n$-th output, to the covariance
perceptron thereby imposes constraints on all $n-1$ other weight
vectors to the already existing outputs.  This causes the classification
problem to become harder the more output covariances have to be tuned.
It explains the decline in pattern capacity by a factor $n-1$ in
Eq.~(\ref{eq:p_final}). Overall, the covariance perceptron has superior
pattern capacity in the case of $n<5$ outputs.

\subsection*{The information capacity exceeds that of the classical perceptron.}

Although the covariance perceptron can classify less patterns than
the classical perceptron in the case of many outputs, one has to take
into account that each pattern has a much higher information content
in the case of the covariance perceptron ($m(m-1)/2$ vs $m$ bits
in the input and $n(n-1)/2$ vs $n$ bits in the output). The information
capacity, that is the amount of bits a classical computer would need
to store a lookup table to implement the same classification as performed
by the covariance perceptron, is (with $K=2^{m(m-1)/2}$, $L=2^{n(n-1)/2}$)
given by
\begin{eqnarray*}
\mathcal{I}^{\mathrm{cov}}(\kappa) & = & \mathcal{P}^{\mathrm{cov}}(\kappa)\left(m(m-1)/2+n(n-1)/2\right).
\end{eqnarray*}
Note that we here, for simplicity, considered the case $f=1$ (which
in general breaks the positive definiteness of the covariance patterns).
The general case $f\neq1$ is discussed in \prettyref{sec:infodensity}.
We notice that the expression for the information capacity of the
covariance perceptron grows $\propto m^{2}(m-1)/(n-1)$, while the
former for the classical perceptron grows with $m^{2}$ (Eq.~\ref{eq:info_class}).

\begin{figure}
\centering{}\includegraphics{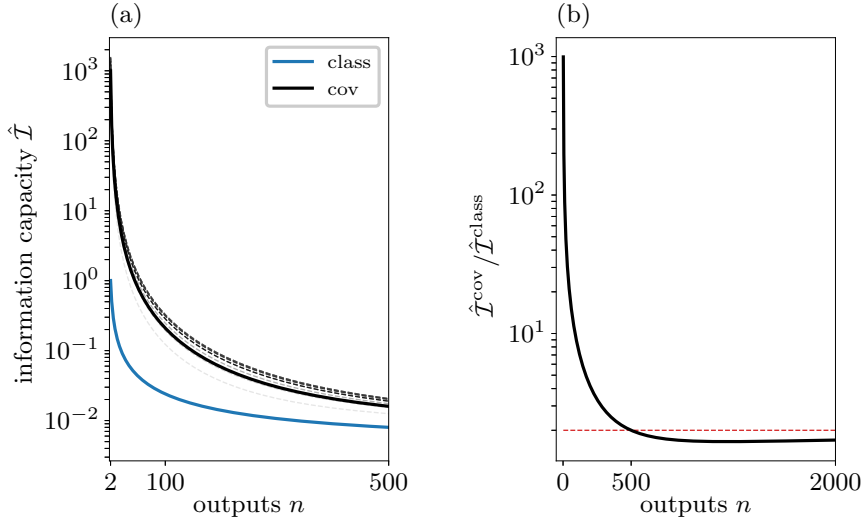}\caption{\textbf{Information capacity.} (\textbf{a}) Information capacity per
synapse as a function of the number of outputs $n$ for the classical
(blue) and covariance (gray shades) perceptron. Solid black curve
for sparsity $f=1$, dashed gray curves for other levels of sparsity
$f\in[0,1]$. (\textbf{b}) Ratio $\hat{\mathcal{I}}^{\mathrm{cov}}(\kappa)/\hat{\mathcal{I}}^{\mathrm{class}}(\kappa)$
given by (\ref{eq:I_cov_over_I_class}) as a function of the number
of outputs $n$. Red dotted horizontal line at ratio $2$. Other parameters:
$m=500$, $\kappa=0$. \label{fig:Info_cap}}
\end{figure}

In the brain, each neuron makes up to thousands of connections. Therefore,
connections are the main objects that cost space. In addition, the
number of synaptic events per time is a common measure for energy
consumption. An important measure for classification performance is,
therefore, the information capacity per synapse $\hat{\mathcal{I}}$
(\prettyref{fig:Info_cap}). For this measure, we get
\begin{eqnarray}
\frac{\hat{\mathcal{I}}^{\mathrm{cov}}(\kappa)}{\hat{\mathcal{I}}^{\mathrm{class}}(\kappa)} & = & \frac{\mathcal{I}^{\mathrm{cov}}(\kappa)}{\mathcal{I}^{\mathrm{class}}(\kappa)}\stackrel{m\gg1}{\approx}\left\{ \begin{array}{cc}
2 & ,n\gg m\\
2 & ,n=m\\
2\frac{m-1}{n-1} & ,n\ll m
\end{array}\right.\label{eq:I_cov_over_I_class}
\end{eqnarray}
For a network with $m=n$, we get $\hat{\mathcal{I}}^{\mathrm{cov}}(\kappa)\approx2\hat{\mathcal{I}}^{\mathrm{class}}(\kappa)$,
i.e. a two times larger information capacity than for the classical
perceptron (\prettyref{fig:Info_cap}b). The same is true if the number
of outputs is much larger than the number of inputs. However, for
networks with strong convergence, i.e. $n\ll m$, the covariance perceptron
outperforms the classical perceptron by a factor $2(m-1)/(n-1)$ that
can be potentially very large (\prettyref{fig:Info_cap}). Note that
a similar result holds for different levels of sparsity (see \prettyref{sec:infodensity},
\prettyref{fig:Info_cap}a). Therefore, in networks that perform a
strong compression of inputs, the covariance perceptron much more
efficiently uses its connections to store information about the stimuli.

\subsection*{The numerical solution is an NP-hard problem}

To check the prediction by the theory, we compare it to numerical
experiments. The problem of finding the weight vectors can be formulated
as maximizing the margin given a certain pattern load. This approach
is the same as for the classical perceptron: Here, the training can
be reduced to a quadratic programming problem \cite[eq. 10.3]{Vapnik98},
the minimization of the length of the readout vector under the constraints
that all patterns be classified with unit margin.

Since the margin $\kappa$ is a non-analytic function due to the appearance
of the minimum operation \prettyref{eq:min_margin}, it cannot be
used directly to perform a gradient descent with regard to the weights.
We thus employ an analytical approximation of the margin, the soft-margin
\[
\kappa_{\eta}:=-\frac{1}{\eta}\,\ln\,\left(\sum_{r}^{p}\sum_{s}^{N}\,\exp\big(-\eta\,\zeta_{s}^{r}G_{s}^{r}\big)\right),
\]
which covaries with $\kappa=\lim_{\eta\rightarrow\infty}\kappa_{\eta}$,
and can be optimized via a standard gradient ascent (see \prettyref{sec:Optimization}).
Physically this soft-margin can be interpreted as a system at finite
inverse temperature $\eta$ for which we find the set of parameters,
the weight vectors, which maximize the free energy $\kappa_{\eta}$.
The states of the system here comprise a discrete set, given by the
patterns $\{\zeta^{r}P^{r}\}_{r}$. In the limit of vanishing temperature,
$\eta\to\infty$, the soft-margin approaches the true margin. It is
also obvious, by Hoelder's inequality, that the soft-margin is convex
in each of the two weight vectors.

\begin{figure}
\centering{}\includegraphics{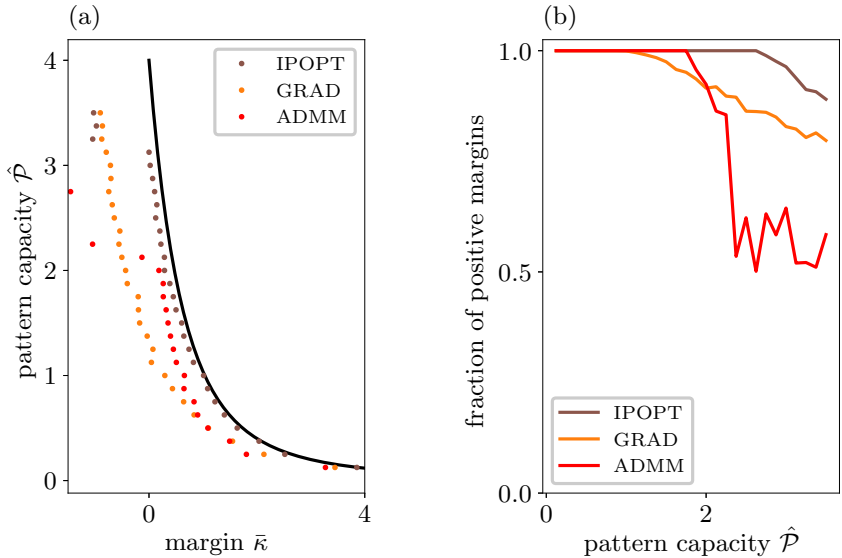}\caption{\textbf{Numerical simulations of pattern capacity.} (\textbf{a}) Pattern
capacity per synapse for a single-readout covariance perceptron as
a function of the margin $\bar{\kappa}$. Symbols show maximum margin
across $N_{\mathrm{init}}$ initial conditions for different numerical
optimizations (brown: IPOPT ($N_{\mathrm{init}}=10$), orange: gradient
ascent of soft-margin ($N_{\mathrm{init}}=10$), red: ADMM ($N_{\mathrm{init}}=5$));
solid curve from theory in the large $m$ limit. (\textbf{b}) Fraction
of positive margins as a function of the pattern capacity per synapse
for different numerical schemes. Parameters: $m=100$, $n=2$, $f=0.2$,
$c=0.5$. \label{fig:capacity}}
\end{figure}

The resulting capacity curve is shown in \prettyref{fig:capacity}a.
We observe that the achieved margin is well below the theoretical
prediction. A better numerical implementation follows from a formulation
as a quadratically constrained quadratic programming problem (cf.
\prettyref{sec:appendix_implementation_QCQP}). For general constraints,
these problems are typically NP-hard. Using an approximate procedure
that iteratively improves an initial guess (alternating directions
method of multipliers (ADMM), \cite{Park17_1703}), yields a margin
that is comparable to the solution obtained by the gradient ascent
for low pattern loads and slightly superior for larger pattern loads,
as shown in \prettyref{fig:capacity}. As the load increases beyond
the point $\hat{\mathcal{P}}\gtrsim2$, the method typically does
not find a solution anymore; a large fraction of patterns have a negative
margin and are thus classified wrongly (\prettyref{fig:capacity}b).
Employing, instead, an interior point optimizer (IPOPT, \cite{Waechter06_25})
yields a significantly larger margin for all pattern loads up to $\hat{\mathcal{P}}\approx3$,
where this scheme breaks down (\prettyref{fig:capacity}b). The result
of the interior point optimizer compares well to the theoretical prediction
in the regime of low pattern load. For larger loads, the numerically
found margin, however, is still slightly smaller than predicted by
the replica-symmetric mean-field theory. Running the optimization
for different initial conditions results in slightly different results
for the margin (only the maximal one is shown in \prettyref{fig:capacity}a)
indicating that the optimizer does not reliably find the unique solution
that is predicted by the theory.

\section{Discussion\label{sec:Discussion}}

In this work, we study information processing of networks that, like
biological neural networks, process temporal signals. We investigate
the scenario where the input-output mapping is dominated by the linear
response, which is a good approximation for small signals in the typical
dynamical regime of cortical networks \cite{Brunel00,Lindner05_061919}.
Focusing only on the temporal mean of these signals, such networks
act as a classical perceptron if a classification threshold is applied
to the outputs. Covariances of small temporal signals, however, transform
in a bilinear fashion in terms of weights, giving rise to what we
call a 'covariance perceptron'. The presented theoretical calculations
focus on the pattern and information capacities of such a covariance
perceptron. The theory uses Gardner's theory of connections ~\cite{Gardner88_257}
in the thermodynamic limit, toward infinitely many inputs ($m\to\infty$).
We have shown that the covariance perceptron indeed presents an analytically
solvable model in this limit and compute the pattern capacity by replica
symmetric mean-field theory, analogous to the classical perceptron~\cite{Gardner88_257}.
It turns out that the pattern capacity exceeds that of the classical
perceptron for a single readout covariance ($n=2$), whereas it decreases
in the case of many outputs. The information capacity in bits of the
covariance perceptron grows with $m^{2}(m-1)/(n-1)$, whereas it only
has a dependence as $m^{2}$ for the classical perceptron. The proposed
paradigm in large and strongly convergent networks can therefore reach
an information capacity that is orders of magnitude higher than that
of the classical perceptron.

The dependence of the pattern capacity on the number of readout neurons
$n$ is a non-trivial result in the case of the covariance perceptron,
because different entries $Q_{ij}$ here share the same rows of the
matrix $W$. These partly confounding constraints reduce the capacity
from the naively expected independence of the $n(n-1)/2$ readouts
to a factor $(n-1)^{-1}$. This factor signifies that with increasing
numbers of readouts, the number of potentially confounding requirements
on the readout vectors rises. Likewise, the capacity cannot simply
be estimated by counting numbers of free parameters.

We demonstrate in \cite{Gilson19_562546} that the paradigm of classification
that we analyzed here can indeed be implemented by means of a network
of linear autoregressive processes. In particular, it is possible
to derive learning rules that are local in time, which tune the readout
vectors for binary classification of covariance patterns. Estimating
covariance patterns from a time series naturally requires the observation
of the process for a certain duration. The resulting estimation noise
can be shown not to be detrimental for the performance of the classification
\cite{Gilson19_562546}. The gradient-based learning rule derived
in \cite{Gilson19_562546} does not reach as superior performance
for single readouts as derived in this manuscript. The reason is twofold:
1. Gardner's theory predicts the theoretical optimum for the capacity,
but it is agnostic to the learning process that should reach this
optimum. Any biophysical implementation of the learning therefore
is likely to yield worse performance. 2. The non-biophysical learning
by gradient-based soft-margin optimization studied here is also incapable
of reaching the theoretical optimum, showing that gradient-based methods
struggle with the non-convexity of the optimization problem. Yet,
the learning rule in \cite{Gilson19_562546} yields as good results
as for the classical perceptron.

The seminal work by Gardner \cite{Gardner88_257} spurred many applications
and extensions. For example, the information capacity within the error
regime \cite{Brunel92_5017} or the computation of the distribution
of synaptic weights of the Purkinje cell \cite{Brunel04_745}. Recently,
the theory has been extended to the classification of data points
that possess a manifold structure \cite{Chung18_031003}. So far,
these works employed a linear mapping prior to the threshold operation.
Here we turn to bilinear mappings in terms of weights and show their
tight relation to the classical perceptron, but expose also striking
differences. The bilinear mapping that we considered arose here from
the mapping of covariance matrices by a linear network dynamics.

The reduction of dimensionality of covariance patterns \textemdash from
many input nodes to a few output nodes\textemdash{} implements an
``information compression''. For the same number of input-output
nodes in the network, the use of covariances instead of means makes
a higher-dimensional space accessible to represent input and output,
which may help in finding a suitable projection for a classification
problem. It is worth noting that applying a classical machine-learning
algorithm, like the multinomial linear regressor~\cite{Bishop06},
to the vectorized covariance matrices corresponds to $nm(m-1)/2$
weights to tune, to be compared with only $nm$ weights in our study
(with $m$ inputs and $n$ outputs). 

The presented calculations are strictly valid only in the thermodynamic
limit $m\rightarrow\infty$. The finite-size simulations that we presented
here agree well, but also show differences to the theory. The discrepancies
between the analytical prediction from the replica-symmetric mean-field
theory and the numerically obtained optimization of the margin may
have different sources. The first are true finite-size effects. Corrections
would technically correspond to taking fluctuations of the auxiliary
fields into account in addition to their saddle point values. A qualitatively
different source of discrepancy arises from the method of training
that we devised here. We first used a gradient ascent of a soft-margin.
The latter only approximately agrees to the true margin. Also, as
the training is implemented by a gradient ascent of the margin, stopping
too early at large system size may lead to an underestimation of the
theoretically possible margin. Analogous to classical perceptron learning,
which, formulated as a support vector machine, can be recast into
a quadratic programming problem \cite{Vapnik98}, the covariance perceptron
maps to a quadratically constrained quadratic programming problem.
These problems are in general NP hard, so that training time increases
exponentially with system size. It is clear that the complexity of
the optimization problem increases with pattern load, because each
pattern contributes one additional inequality constraint. The observed
deviations appear at a pattern load of $\mathcal{P}\gtrsim3$. It
is likely that they are due to the inability of the numerical solver
to identify the unique optimal solution.

Another possibility is that indeed multiple solutions with similar
margins exist if the load exceeds a certain point. This situation
would correspond to replica symmetry breaking, because the existence
of multiple degenerate solutions for the readout vector would show
up as different replica settling in either of these solutions; analogous
to a disordered system, which possesses a number of nearly degenerate
meta-stable ground states \cite{Fischer91}. In Gardner's theory we
search for the point at which the overlap between replica becomes
unity; this assumption would have to be relaxed. It is conceivable
that instead the set of these solutions vanishes together as the pattern
load is increased beyond the capacity limit. Future work should address
this question, either by the application of more powerful numerical
optimizers or by analyzing the replica-symmetric mean-field theory
with regard to instabilities of the symmetric solution.

The analysis presented here assumed the classification of uncorrelated
patterns. In applications, however, the data to be classified typically
has some internal structure. In the simplest case, different patterns
could be correlated with correlations of a certain order. Second order
correlations between patterns, for example, would show up in the pattern
average \prettyref{eq:pattern_average} as additional quadratic terms,
which would require the introduction of additional auxiliary fields
for decoupling. A detailed analysis of this extension is left for
future work.

The approximation of the network mapping in linear response theory
here leads to a straight forward relationship between input and output
covariances that is bilinear in the feed-forward connectivity matrix.
In recurrent networks, linear response theory would amount to determining
the network propagator. Such mappings are of the form $W(\omega)=(1+H(\omega)\,J)^{-1}$,
where $H(\omega)$ is the Fourier transform of the temporal linear
response kernel of a neuron and $J$ the recurrent connectivity. This
expression shows that the mapping between input and output covariances
becomes more involved in the recurrent setting. Analyzing this situation
may be possible by expressing the matrix inversion that appears in
the propagator by help of a Gaussian integral. We leave this problem
open for future work. An alternative approach first determines the
optimal bilinear readout, as presented here, and subsequently determines
the parameters of the recurrent network, foremost its connectivity,
to implement this mapping.

Another extension consists in considering patterns of higher-than-second-order
correlations. Generalizing the obtained results, higher-dimensionality
may lead to higher information capacity when large number of inputs
are considered. In contrast, this also implies additional constraints
that limit the information capacity as shown when increasing the number
of outputs for the present case of second-order correlations. There
should thus be a trade-off for optimal information capacity that depends
on the correlation order and the number of inputs and outputs.

Going beyond linear response theory is another route that may lead
to a problem of similar structure as the system studied here. Concretely,
assume the simplest setting of a static input-output mapping described
by a quadratic gain function $y=f(z)=z^{2}$ of a neuron. When mapping
the summed synaptic input $z_{i}=\sum_{k}w_{ik}x_{k}$, the mean of
the output $Y_{i}=\big[W\tilde{P}W^{\T}\big]_{ii}$ contains a bilinear
term in $W$ that maps the matrix of second moments $\tilde{P}$ of
the data to the first moment of the network's output. This mapping
is similar to the one studied here. A difference is, though, the dimension
of the output, which in the latter case equals the number of units.
Analyzing this system may be an interesting route for future studies.

A central motivation to study the current setting comes from the need
to derive a self-consistent theory of biological information processing.
As learning at the synaptic level is implemented by covariance-sensitive
learning rules, using a covariance-based representation of information
thus allows the construction of a scheme that is consistent at all
levels. A further crucial ingredient would be the study of the covariance
mapping across multiple layers of processing and including the abundant
presence of recurrence.

\ack{}{}

This work was partially supported by the Marie Sklodowska-Curie Action
(Grant H2020-MSCA-656547) of the European Commission, the Helmholtz
young investigator's group VH-NG-1028, the European Union's Horizon
2020 research and innovation programme under grant agreement No.\ 785907
(Human Brain Project SGA2), the Exploratory Research Space (ERS) seed
fund neuroIC002 (part of the DFG excellence initiative) of the RWTH
university and the JARA Center for Doctoral studies within the graduate
School for Simulation and Data Science (SSD).

\section{Appendix\label{sec:Appendix}}

\subsection{Optimization of the soft-margin\label{sec:Optimization}}

In order to numerically test the theoretical predictions, we need
to maximize Eq.~(\ref{eq:min_margin}). A gradient-based optimization
is, however, unfeasible due to the non-analytical minimum operation.
Therefore, as an alternative, we consider the following soft-minimum
as an objective function 
\begin{eqnarray}
O(W) & := & -\kappa_{\eta}(W)=\frac{1}{\eta}\,\ln\,\left(\sum_{r=1}^{p}\,\sum_{i<j}\,\exp\big(-\eta\,\zeta_{ij}^{r}(WP^{r}W^{\T})_{ij}\big)\right)\label{eq:soft_margin}\\
 & = & \frac{1}{\eta}\,\ln\,\left(\sum_{r=1}^{p}\,\frac{1}{2}\sum_{i\neq j}\,\exp\big(-\eta\,\zeta_{ij}^{r}(WP^{r}W^{\T})_{ij}\big)\right).\nonumber 
\end{eqnarray}

This definition has the form of a scaled (scaling parameter $\eta$)
cumulant-generating function, if we consider the patterns to be drawn
randomly. In the limit $\eta\to\infty$, this objective function will
be dominated by the points with the smallest margins, so we recover
\begin{equation}
\kappa_{\eta}(W)\stackrel{\eta\to\infty}{\to}\min_{ij,r}\,\big(\zeta_{ij}^{r}(WP^{r}W^{\T})_{ij}\big).\label{eq:def_margin}
\end{equation}
We use this objective function $O(W)$ with finite $\eta$: Larger
$\eta$ causes stronger contribution of patterns classified with small
margin.

\begin{figure}
\centering{}\includegraphics{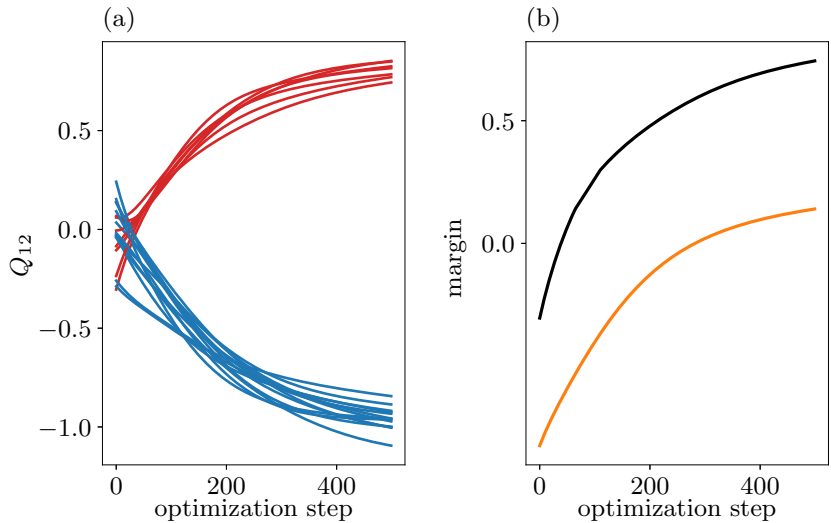}\caption{\textbf{Numerical optimization of soft-margin.} (\textbf{a}) Evolution
of the readout covariances $Q_{12}^{r}=(W\,P^{r}\,W^{\protect\transp})_{12}$
in the optimization by a gradient ascent based on the soft-margin
$\kappa_{\eta}$ in place of $\kappa$. Each curve corresponds to
one of the $p=20$ input patterns: red for $\zeta_{12}^{r}=1$ and
blue for $\zeta_{12}^{r}=-1$. (\textbf{b}) Coevolution of the margin
$\kappa$ (black curve) and the soft-margin $\kappa_{\eta}$ (orange
curve). Parameters: $m=100$, $f=0.2$, $c=0.5$, $\eta=4$, $\iota=0.01$.\label{fig:optimization}}
\end{figure}

 In order to check the analytical prediction for the maximum pattern
capacity $\mathcal{P}$ (\prettyref{fig:capacity}), we need to find
\begin{equation}
W_{\mathrm{opt}}=\arg\min_{W}O(W).\label{eq:objective}
\end{equation}
We solve the minimization problem by gradient descent, which yields,
with $\zeta_{ij}^{r}(WP^{r}W^{\T})_{ij}=\zeta_{ij}^{r}\,\sum_{kl}^{m}W_{ik}P_{kl}^{r}W_{jl}$,
the gradient
\begin{eqnarray*}
\frac{\partial O(W)}{\partial W_{uv}} & = & \frac{1}{\eta}\,\frac{\sum_{r}^{p}\,\frac{1}{2}\sum_{i\neq j}\,\frac{\partial A_{ij}^{r}}{\partial W_{uv}}}{\sum_{r}^{p}\,\frac{1}{2}\sum_{i\neq j}\,A_{ij}^{r}},\\
A_{ij}^{r} & = & \exp\big(-\eta\,\zeta_{ij}^{r}(WP^{r}W^{\T})_{ij}\big),\\
\frac{\partial A_{ij}^{r}}{\partial W_{uv}} & = & (-\eta\,\zeta_{ij}^{r})\,(\delta_{iu}\,\sum_{l}^{m}P_{vl}^{r}W_{jl}+\delta_{ju}\,\sum_{k}^{m}W_{ik}P_{kv}^{r})\,A_{ij}^{r}.
\end{eqnarray*}
We update the readout vectors by
\begin{eqnarray*}
\Delta W_{uv} & = & -\iota\,\frac{\partial O(W)}{\partial W_{uv}},
\end{eqnarray*}
where $\iota>0$ is the learning rate, here set to be $\iota=0.01$.
The normalization of the readout vectors is taken care of by enforcing
unit length after each learning step (\prettyref{fig:optimization}).

\subsection{Formulation as quadratically-constrained quadratic programming problem\label{sec:appendix_implementation_QCQP}}

To obtain a numerical solution for the covariance perceptron, it is
convenient to reformulate the optimization problem in the form of
a quadratically constrained quadratic programming problem. These problems
frequently occur in different fields of science and efficient numerical
solvers exist. The idea is analogous to the formulation of the support
vector machine learning in terms of a quadratic programming problem
\cite[eqs. (10.3) and (10.4)]{Vapnik98}. For $n=2$, an equivalent
problem to finding the bi-linear readout with unit-length readout
vectors $w_{i}\in\bR^{n}$, $i=1,2$ that maximize the margin, is
the optimization problem
\begin{eqnarray*}
\text{minimize}: &  & ||\tilde{w}_{1}||^{2}\\
\\
\text{with constraints:}\\
 &  & ||\tilde{w}_{1}||^{2}=||\tilde{w}_{2}||^{2},\\
 &  & \zeta_{12}^{r}\tilde{w}_{1}^{\T}P^{r}\tilde{w}_{2}\ge1\quad\forall\,1\le r\le p.
\end{eqnarray*}
The constraints can be formulated conveniently by combining the pair
of readout vectors into a single vector $v=(\tilde{w}_{1},\tilde{w}_{2})\in\bR^{2n}$
and defining for the length constraint a matrix 
\begin{eqnarray*}
A_{ij} & := & \left\{ \begin{array}{cc}
\delta_{ij} & 1\le i\le n\\
-\delta_{ij} & n<i\le2n
\end{array}\right.,
\end{eqnarray*}
so that a pair of quadratic form inequality constraints $v^{\T}Av\ge0$
and $v^{\T}Av\le0$ fixes the length of the two readout vectors to
be identical. Analogously one defines for each pattern $r$ one matrix
and a bi-linear inequality constraint to enforce a margin of at least
unity as
\begin{eqnarray*}
B_{ij}^{r} & := & \frac{1}{2}\left(\begin{array}{cc}
0 & \zeta^{r}P^{r}\\
\zeta^{r}(P^{r})^{\T} & 0
\end{array}\right),\\
v^{\T}Bv & \ge & 1.
\end{eqnarray*}

The transposition of the matrix $(P^{r})^{\T}$ appearing in the lower
left element, for symmetric matrices (covariances), can of course
be left out. But the formulation also holds for non-symmetric matrices.
The task is thus to minimize the norm of $v$ under $p+2$ quadratic
inequality constraints. Random vectors with independent Gaussian entries,
normalized to unit length, serve as initial guess. The normalized
readout vectors are finally obtained as $w_{i}=\tilde{w}_{i}/||\tilde{w}_{i}||$.

We here use an interior point method implemented in the package IPOPT
\cite{Waechter06_25}, with a frontend provided by the python package
QCQP \cite{Park17_1703} within the domain-specific language CVXPY
\cite{Diamond16_1}. 

\subsection{Information capacity for sparse patterns\label{sec:infodensity}}

The number of configurations of a sparse covariance pattern is
\begin{equation}
K=\left(\begin{array}{c}
M\\
fM
\end{array}\right)2^{fM},
\end{equation}
where $\left(\begin{array}{c}
M\\
fM
\end{array}\right)$ is the number of ways to position $fM$ non-zero entries in a single
pattern of length $M$, and $2^{fM}$ is the number of configurations
of the non-zero entries in a single pattern. The information capacity
follows as
\begin{eqnarray*}
\mathcal{I}^{\mathrm{cov}}(\kappa) & = & \mathcal{P}^{\mathrm{cov}}(\kappa)\left[fM+\log_{2}\left(\begin{array}{c}
M\\
fM
\end{array}\right)+N\right]\\
 & \approx & \mathcal{P}^{\mathrm{cov}}(\kappa)\left[M(f-S(f))+N\right]
\end{eqnarray*}
where we used $\log_{2}\left(\begin{array}{c}
M\\
fM
\end{array}\right)=-MS(f)$ with $S(f)=\left(f\log_{2}(f)+(1-f)\log_{2}(1-f)\right)$. Although
the calculations in \prettyref{sec:Theory} ignore the constraint
that the covariance matrices $P^{r}$ must be positive semidefinite,
this constraint is ensured when using not too dense and strong entries
such that $fc\ll1$, thanks to the unit diagonal. Since $\sqrt{fc^{2}}$
only determines the scale on which the margin $\kappa$ is measured,
the optimal capacity can always be achieved if one allows for a sufficiently
small margin. As shown in \prettyref{fig:Info_cap}b, the level of
sparsity only has a minor impact on the information capacity per synapse
when comparing to the classical perceptron.

\providecommand{\newblock}{} 
\end{document}